\documentclass[letterpaper,twocolumn,prl,aps,superscriptaddress,floatfix]{revtex4}
\usepackage{mathptmx}

\usepackage{amsmath}
\usepackage{amssymb}
\usepackage{graphicx}
\usepackage{esint}
\usepackage{float}
\usepackage{url}
\usepackage{mhchem}

\makeatletter

\pdfpageheight\paperheight
\pdfpagewidth\paperwidth

\@ifundefined{textcolor}{}
{%
 \definecolor{BLACK}{gray}{0}
 \definecolor{WHITE}{gray}{1}
 \definecolor{RED}{rgb}{1,0,0}
 \definecolor{GREEN}{rgb}{0,1,0}
 \definecolor{BLUE}{rgb}{0,0,1}
 \definecolor{CYAN}{cmyk}{1,0,0,0}
 \definecolor{MAGENTA}{cmyk}{0,1,0,0}
 \definecolor{YELLOW}{cmyk}{0,0,1,0}
}

\usepackage{xcolor}\usepackage{soul}

\definecolor{blue}{rgb}{0,0,1}
\definecolor{red}{rgb}{1,0,0}
\definecolor{green}{rgb}{0,1,0}

\usepackage[unicode=true,pdfusetitle,
 bookmarks=true,bookmarksnumbered=false,bookmarksopen=false,
 breaklinks=false,pdfborder={0 0 0},pdfborderstyle={},backref=false,colorlinks=true]
 {hyperref}
\hypersetup{
 colorlinks,linkcolor=red,citecolor=blue}
\makeatother

\begin{document}



\title{Vacuum-gap transmon qubits realized using  flip-chip technology}

\author{Xuegang Li}
\thanks{These authors contributed equally.}
\affiliation{%
	Beijing Academy of Quantum Information Sciences,  Beijing 100193, China
}%
\author{Yingshan Zhang}
\thanks{These authors contributed equally.}
\affiliation{%
	Beijing Academy of Quantum Information Sciences, Beijing 100193, China
}%
\author{Chuhong Yang}
\thanks{These authors contributed equally.}
\affiliation{%
	Beijing Academy of Quantum Information Sciences,  Beijing 100193, China
}%
\author{Zhiyuan Li}
\affiliation{%
	Beijing Academy of Quantum Information Sciences,  Beijing 100193, China
}%
\author{Junhua Wang}
\affiliation{%
	Beijing Academy of Quantum Information Sciences,  Beijing 100193, China
}%
\author{Tang Su}
\affiliation{%
	Beijing Academy of Quantum Information Sciences,  Beijing 100193, China
}%
\author{Mo Chen}
\affiliation{%
	Beijing Academy of Quantum Information Sciences,  Beijing 100193, China
}%
\author{Yongchao Li}
\affiliation{%
	Beijing Academy of Quantum Information Sciences,  Beijing 100193, China
}%
\author{Chengyao Li}
\affiliation{%
	Beijing Academy of Quantum Information Sciences,  Beijing 100193, China
}%
\author{Zhenyu Mi}
\affiliation{%
	Beijing Academy of Quantum Information Sciences,  Beijing 100193, China
}%
\author{Xuehui Liang}
\affiliation{%
	Beijing Academy of Quantum Information Sciences,  Beijing 100193, China
}%
\author{Chenlu Wang}
\affiliation{%
	Beijing Academy of Quantum Information Sciences,  Beijing 100193, China
}%
\author{Zhen Yang}
\affiliation{%
	Beijing Academy of Quantum Information Sciences,  Beijing 100193, China
}%
\author{Yulong Feng}
\affiliation{%
	Beijing Academy of Quantum Information Sciences,  Beijing 100193, China
}%
\author{Kehuan Linghu}
\affiliation{%
	Beijing Academy of Quantum Information Sciences,  Beijing 100193, China
}%
\author{Huikai Xu}
\affiliation{%
	Beijing Academy of Quantum Information Sciences,  Beijing 100193, China
}%
\author{Jiaxiu Han}
\affiliation{%
	Beijing Academy of Quantum Information Sciences,  Beijing 100193, China
}%
\author{Weiyang Liu}
\affiliation{%
	Beijing Academy of Quantum Information Sciences,  Beijing 100193, China
}%
\author{Peng Zhao}
\affiliation{%
	Beijing Academy of Quantum Information Sciences,  Beijing 100193, China
}%
\author{Teng Ma}
\affiliation{%
	Beijing Academy of Quantum Information Sciences,  Beijing 100193, China
}%
\author{Ruixia Wang}
\affiliation{%
	Beijing Academy of Quantum Information Sciences,  Beijing 100193, China
}%
\author{Jingning Zhang}
\affiliation{%
	Beijing Academy of Quantum Information Sciences,  Beijing 100193, China
}%
\author{Yu Song}
\affiliation{%
	Beijing Academy of Quantum Information Sciences,  Beijing 100193, China
}%
\author{Pei Liu}
\affiliation{State Key Laboratory of Low Dimensional Quantum Physics, Department of Physics, Tsinghua University, Beijing 100084, China}
\author{Ziting Wang}
\affiliation{Beijing National Laboratory for Condensed Matter Physics,
	Institute of Physics, Chinese Academy of Sciences, Beijing 100190, China}
\author{Zhaohua Yang}
\affiliation{Beijing National Laboratory for Condensed Matter Physics,
	Institute of Physics, Chinese Academy of Sciences, Beijing 100190, China}
\author{Guangming Xue}
 \email{xuegm@baqis.ac.cn}
\affiliation{%
	Beijing Academy of Quantum Information Sciences,  Beijing 100193, China
}%
\author{Yirong Jin}
 \email{jinyr@baqis.ac.cn}
\affiliation{%
	Beijing Academy of Quantum Information Sciences,  Beijing 100193, China
}%
\author{Haifeng Yu}
 \email{hfyu@baqis.ac.cn}
\affiliation{%
	Beijing Academy of Quantum Information Sciences,  Beijing 100193, China
}%


\date{\today}

\begin{abstract}
Significant progress has been made in building large-scale superconducting quantum processors based on flip-chip technology. In this work, we use the flip-chip technology to realize a modified transmon qubit, donated as the "flipmon", whose large shunt capacitor is replaced by a vacuum-gap parallel plate capacitor. To further reduce the qubit footprint, we place one of the qubit pads and a single Josephson junction on the bottom chip and the other pad on the top chip which is galvanically connected with the single Josephson junction through an indium bump. The electric field participation ratio can arrive at nearly 53\% in air when the vacuum-gap is about 5 $\mathrm{\mu m}$, and thus potentially leading to a lower dielectric loss. The coherence times of the flipmons are measured in the range of 30-60\,$\mathrm{\mu}$s, which are comparable with that of traditional transmons with similar fabrication processes. The electric field simulation indicates that the metal-air interface's participation ratio increases significantly and may dominate the qubit's decoherence. This suggests that more careful surface treatment needs to be considered. No evidence shows that the indium bumps inside the flipmons cause significant decoherence. With well-designed geometry and good surface treatment, the coherence of the flipmons can be further improved.

\end{abstract}

\maketitle


\section{\label{sec:intro}introduction}
Superconducting quantum processors have reached the scale of over 50 qubits with high gate fidelity, good addressability~\cite{Google2019,Gongeabg7812}. One of the short-term goals for the quantum computing hardware community is to further increase the scale to the order of 1000 qubits, which may implement some practical applications, such as quantum computational chemistry~\cite{McArdle2020}, combinatorial optimization problem~\cite{harrigan2021quantum}. In the challenge of achieving this goal, qubits with compact design, flexible connectivity, and good coherence are crucial issues. Transmon is one of the most promising candidates to meet these requirements~\cite{Koch2007Charge}. Traditional planar transmon designs face wiring problem, which limits their connectivity. In order to solve this problem, 3D or semi-3D interconnection technologies are intensively explored, including air-bridge~\cite{chen2014fabrication,dunsworth2018method}, flip-chip ~\cite{Foxen_2017,rosenberg20173d,satzinger2019simple}, through-silicon via (TSV)~\cite{yost2020solid,mallek2021fabrication}, pass through holes~\cite{Gongeabg7812}.
Currently, flip-chip technology is the most popular one due to its simple fabrication process and process stability. Besides, flip-chip technology provides another probability for the design of superconducting quantum processors (such as resonators~\cite{satzinger2019simple,kelly2020low}, capacitive couplers~\cite{gold2021entanglement}, and so on).


\begin{figure*}[ht]
	\includegraphics{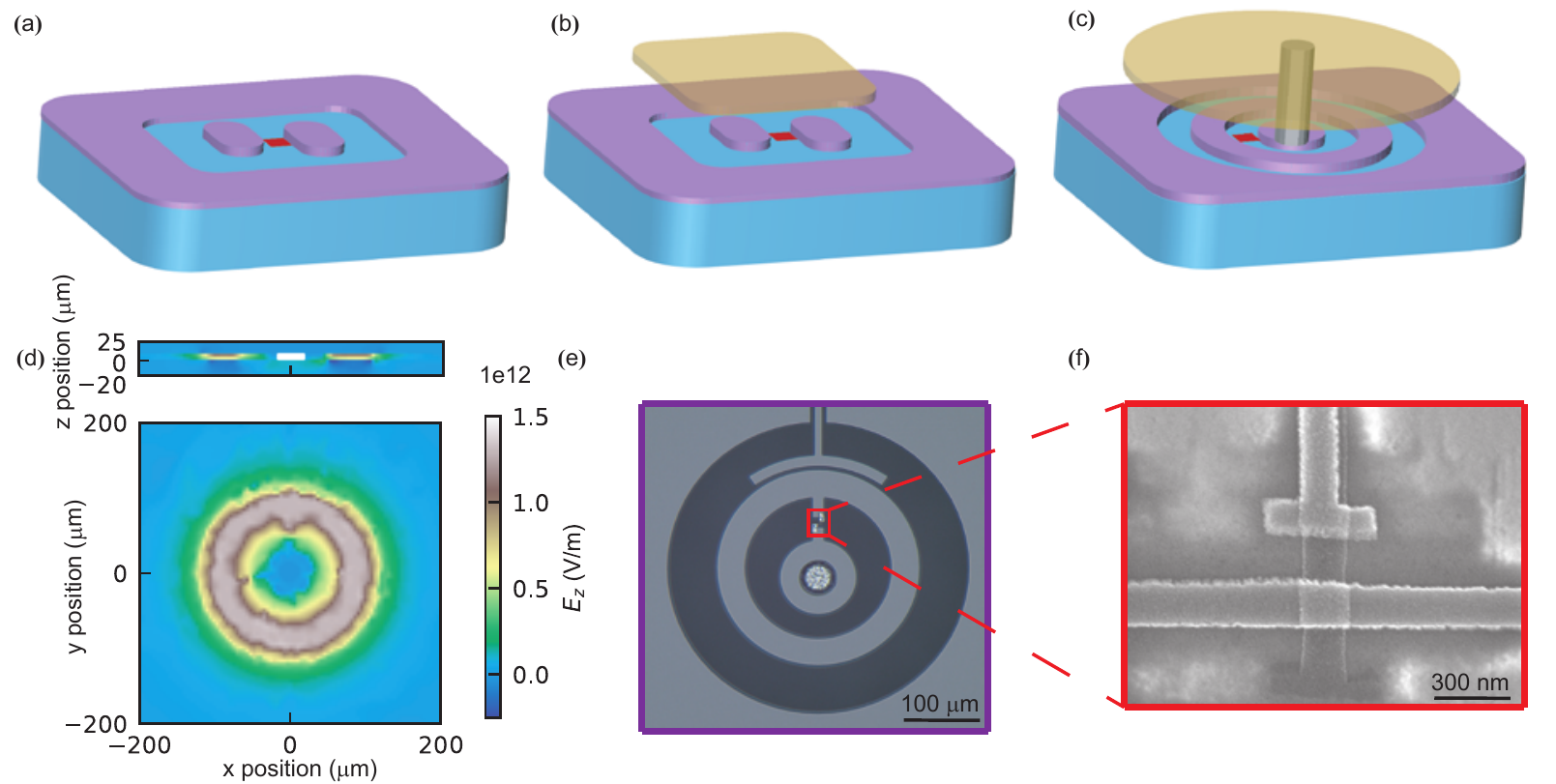}
	\caption{\label{fig:design} Schematics of (a) the traditional transmon, (b) modified transmon with flip-chip, and (c) flipmon. The metal film on the bottom chip (the chip) and the top chip (the carrier) are magenta and yellow, respectively. The single Josephson junction (red) is deposited between two floated capacitor pads on the bottom substrate (blue). The substrate of the chip is not drawn for concision. The metal pad on the chip is either capacitively (b) or galvanically (c) connected with one of the metal pads of the shunt capacitor on the carrier. The vacuum gap between the chip and the carrier is about 5 $\mu m$ in the experiment. (d) The simulated distribution of electrical field in flipmon. Top: side view at the y position = 0 $\mu m$. Bottom: top view at the z position = 5 $\mu m$. The white rectangle in the side view shows that there is no electric field distribution inside the indium bump. (e) Magnified optical micrograph of the geometric structure on the carrier of flipmon. (f) The SEM image of a "T-shaped" junction.}
\end{figure*}

A large shunt capacitor is the key component of the transmon, which can suppress the charging noise and maintain a sufficient anharmonicity. In addition to the traditional transmon with a large coplanar shunt capacitor, a new transmon design (mergemon) which merges the shunt capacitor into the Josephson junction, has been demonstrated in Ref.~\cite{zhao2020merged,mamin2021merged}.  Such a design can achieve a much smaller qubit footprint, whereas a large junction may increase the probability of two-level systems (TLS) in the junction area.
In this work, we propose a new design of transmon by using the flip-chip technology, which we denoted as "flipmon". The central idea of this design is to use a vacuum-gap parallel plate capacitor as the shunt capacitor of transmon. Although vacuum-gap capacitors have been demonstrated in Ref.~\cite{bosman2017multi,cicak2009vacuum}, as far as we know, it is the first time that the vacuum-gap capacitor is used in a transmon design by using the flip-chip technology. This design not only reduces the footprint of the qubit but also has flexible connectivity. The flipmons also increase the electric field participation ratio in the air to nearly 53\%. In contrast, the traditional transmon design has an electric field participation ratio of only about 10\% in the air~\cite{wang2015surface}. The higher electric field participation ratio in the air may decrease the dielectric loss~\cite{cicak2009vacuum}.

We first tested the superconductivity of the flip-chip bumps whose superconducting critical current was over $10~\mathrm{mA}$ with almost 100\% yield. The heights of indium bumps were measured using the scanning electron microscopic (SEM) with deviations less than 0.5 $\mu m$. In order to demonstrate the performance of the flipmons, we measured three flipmon samples with the averaged coherence times in a range of 30-60~$\mathrm{\mu s}$. The detailed simulation of the electric field distribution of the flipmon indicates that the metal-air (MA) interface may be a dominant source of decoherence. We believe that the indium bumps inside the flipmons have little impact on their energy relaxation, as discussed in Section~\ref{sec:meas}. In general, flipmon may be a promising candidate for future high-density large-scale superconducting quantum processors.

\section{\label{sec:design}Flipmon design}
A traditional planar transmon with a floating shunted capacitor is shown in Fig.~\ref{fig:design} (a). Generally, larger capacitor pads lead to a smaller participation ratio of some lossy interfaces, such as the MA, metal-substrate (MS), and substrate-air (SA) interface. As a result, coherence of the qubit can be improved. When scaling up the quantum circuits with transmons, some difficulties, including wiring, crosstalk, etc., appear. Flip-chip technology is a feasible measure to alleviate this problem by bonding two chips together, where indium bumps are commonly used because of their ease of fabrication, soft material properties, and good superconductivity. Here after, for convenience of description, we call the top chip as "chip", and the bottom one as "carrier".





We designed two different qubit schemes using the flip-chip technology to explore the possibility of utilizing the vacuum-gap capacitor in a transmon. The first design is shown in Fig.~\ref{fig:design} (b). It is similar to the traditional planar transmon, except for an additional pad arranged above the capacitor pads on the carrier. The metallizations on the carrier and the chip are represented in magenta and yellow, respectively. The substrate of the chip is not shown for clarity. The second design, denoted as "flipmon", is shown in Fig.~\ref{fig:design} (c). It utilizes the flip-chip technology to place one of the capacitor pads on the chip and the other on the carrier. A single Josephson junction (marked as the red square) formed on the carrier is connected to the chip through an indium bump. This geometry can achieve a smaller qubit footprint.

\begin{table}[t]
\begin{center}
\begin{tabular}{ |c|c| c| c|c| }
    \hline
	 &${\rm Sub_t}$&${\rm SM_t}$&${\rm SA_t}$&${\rm MA_t}$\\
	 \hline
	 $p_i$&0.105&1.31e-5&1.12e-5&3.32e-5\\
	 \hline
	 \hline
	 & Vacuum gap&Indium pump&& \\
	 \hline
	 $p_i$&0.532&1.03e-11&&\\
	 \hline
	 \hline
	 &${\rm Sub_b}$&${\rm SM_b}$&${\rm SA_b}$& ${\rm MA_b}$\\
	 \hline
     $p_i$& 0.363&3.86e-5&1.20e-4&2.07e-5\\
     \hline

\end{tabular}
\end{center}
\label{tab:EPR}
\caption{Energy participation ratio of all components of the flipmon geometry. The subscript ${\rm t}$ in the first row is representing the top chip (the chip). The subscript ${\rm b}$ in the last row is representing the bottom chip (the carrier). The energy participation ratio of the indium pump surface is almost zero and thus leads nearly no impact on the qubit energy relaxation. Over half of the energy is distributed in the vacuum gap and thus can reduce the dielectric loss.}
\end{table}

One of the challenges of the flipmon is whether the indium pump brings additional loss. Fortunately, through the finite element simulation of the electric field distribution, shown in Fig.~\ref{fig:design} (d), we see that there is little electric field around the indium pump. Most of the electric field is concentrated between the parallel plate capacitor, and almost no electric field is distributed on the edge of the pads. As a result, we can ignore the corner effect.  In order to further determine the energy participation ratio of each component, including the dielectric layers and the interface layers, we calculated the fraction of the electric field energy stored in each component relative to the total charge energy,
\begin{equation*}
p_i\approx \int_{V_i} \frac{\epsilon_i \overrightarrow{E_i} \cdot \overrightarrow{E_i^*}}{U_{tot}} ,
\end{equation*}\label{Participation ratio}where $\epsilon_i$ is the dielectric constant of each component and $U_{tot}$ is the energy integral of all objects. The thickness of the interface layer is typical $\sim 3\,nm$, which is challenging to simulate~\cite{wang2015surface}. In addition, we assume a perfect metal film with zero thickness in the simulation and then calculate the corresponding interface electric field according to the boundary condition~\cite{gambetta2016investigating},
\begin{center}
\begin{equation*}
\begin{aligned}
\overrightarrow{E_s}_{\|}=\overrightarrow{E_j}_{\|},\\
\epsilon_s\overrightarrow{E_s}_{\perp}=\epsilon_j\overrightarrow{E_j}_{\perp}
\end{aligned}
\end{equation*}
\end{center}
where $E_s$ is the electric field of each interface component, and $E_j$ is the electric field of the corresponding adjacent dielectric component. For simulation, we set the dielectric constant of all metal oxides to $10$, the thickness of all interface layers to $3\,nm$~\cite{wenner2011surface}, and the vacuum gap between two capacitor pads to $5\,\mathrm{\mu m}$. Then, we obtained the energy participation ratio of all the components, shown in Table I. We can see that the participation ratio of the MA interface is significantly higher than that of the traditional planar transmon.

A full circuit arranged with four flipmons is shown in Fig.~\ref{fig:sample}. For each flipmon, a control line and a readout resonator are coupled to it capacitively. All the readout resonators then coupled to a Purcell filter. The Purcell filter is designed as a $\lambda/2-$type transmission line resonator with asymmetry coupling to the input and output lines through inter-digital capacitors. the coupling capacitor to the output side is much larger than that to the input side. Such a design guarantees that almost all the photons carrying signal goes into the output amplification chain and be measured. To eliminate cross-effects among the qubits, the flipmons are separated far enough without any direct coupling. Furthermore, we use the fixed-frequency qubits in all designs to avoid the influence of the noise from magnetic flux fluctuations. Indium bumps (small circles shown in Fig.~\ref{fig:design}) outside the flipmons are used to connect the ground planes of the chip and the carrier together, so as to suppress the cross-talk of the qubits, similar to the function of air-bridges.

\begin{figure}[t]
	\includegraphics{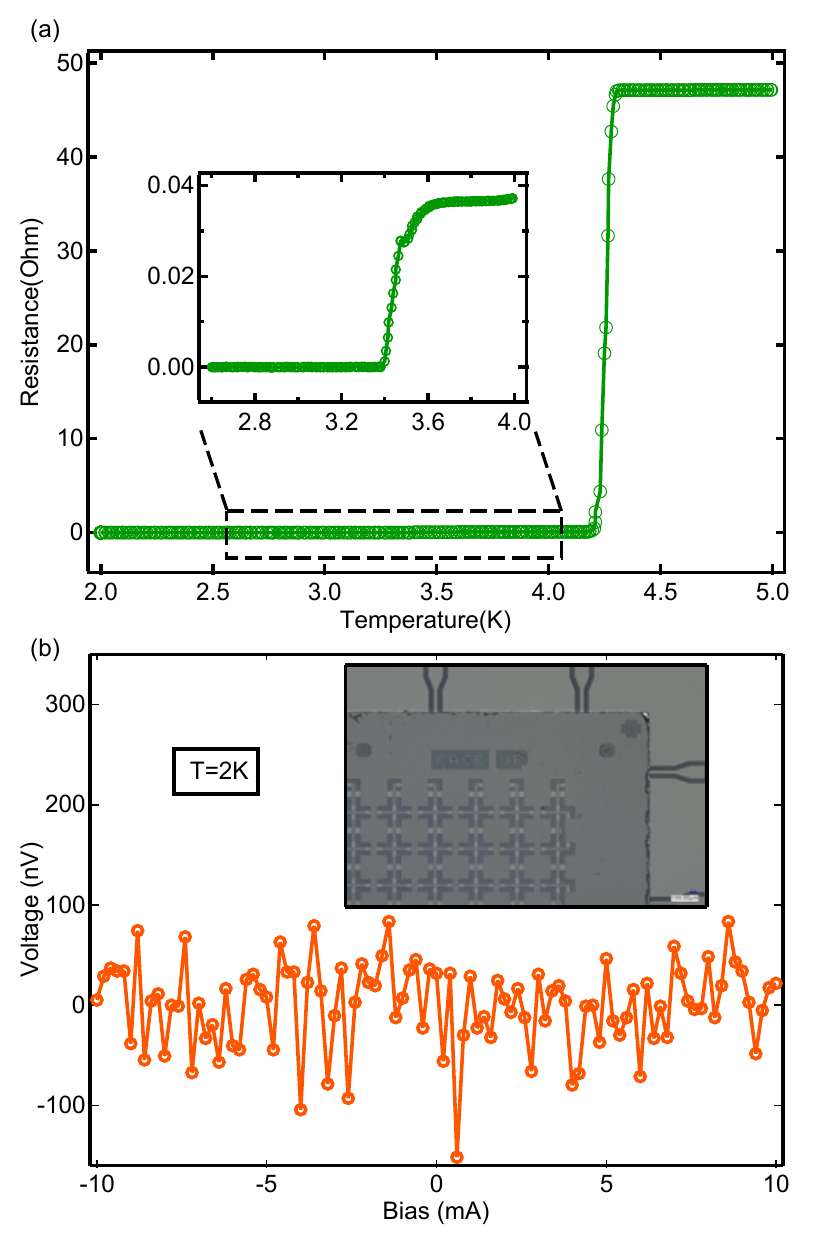}
	\caption{\label{fig:fab1} Characterization of indium bump on tantalum. (a) The R-T curve of a test chip containing \ce{Ta} electrodes and 126 indium bumps with a diameter of 30\,$\mathrm{\mu}$m and height of $\sim 5\, \mu m$. The superconducting transitions at 4.26~K and 3.44~K are due to the tantalum in $\alpha$ phase, and the indium, respectively. (b) The I-V curve of the same chip shows that the critical current of all zero-resistance indium bumps exceeds 10~mA. This is much larger than the critical current of the Josephson junction. The inset is the optical micrograph of a test chip.}
\end{figure}

\begin{figure}[ht]
	\includegraphics{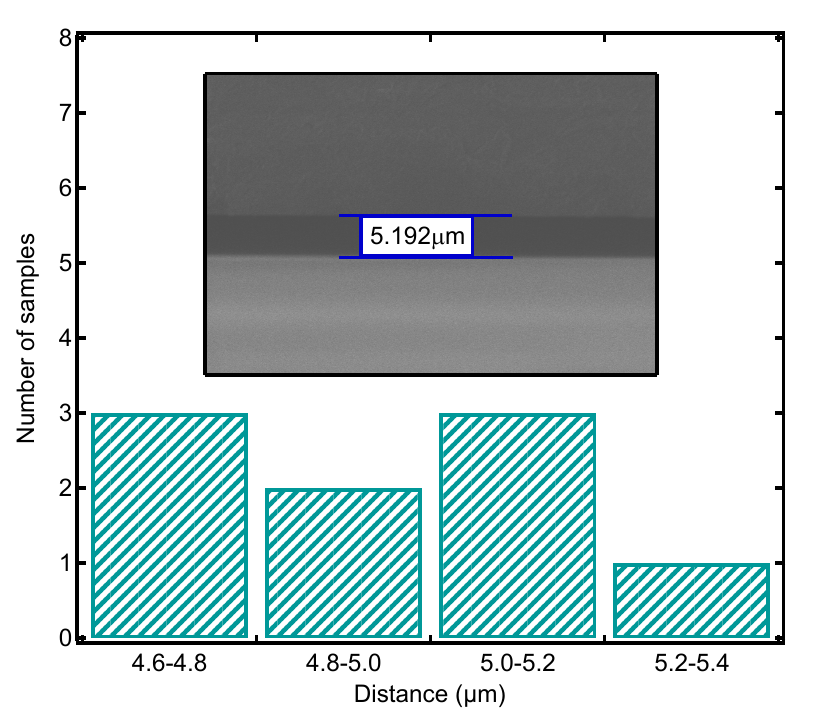}
	\caption{\label{fig:gap} The vacuum gap between the chip and the carrier is determined by the height of the indium bumps. An SEM image of one of the samples is shown in the inset. The average vacuum gap of the nine samples is about $5~\mu m$ with deviations less than $0.5~\mu m$. This stable flip-chip process is attributed to the stop bumps.}
\end{figure}

\section{\label{sec:fab}Flip-chip interconnection and Device Fabrication}
In order to obtain high-quality flip-chip capacitors, first of all, we must get high-quality indium bump flip-chip interconnects with good superconductivity and mechanical properties. We use tantalum instead of aluminum as the superconducting film. First, a significant improvement of coherence time has been achieved. $T_1$ of 0.36\,ms \cite{place2021new} and 0.5\,ms \cite{yu2021tatransmon} has been measured on traditional transmon by resorting to tantalum as the electrodes. Second, aluminum can form a lossy intermetallic layer with indium, while tantalum is a hard metal, and it cannot form a stable phase of TaIn alloy at low temperature\cite{atomly}. As a result, we did not add any under bump metallization (UBM) layer during the indium bump fabrication process. Indium bumps were directly grown on the tantalum film at speed exceeding 2 nm/sec through a thermal evaporation system. Before evaporation, the surface of the films was cleaned by an ion milling process to increase adhesion~\cite{schulte2012characterization,kim2008effect} and to ensure that indium and tantalum formed a superconducting connection. Approximately 5 $\mathrm{\mu m}$'s indium bumps with a diameter of 20 to 50 $\mathrm{\mu m}$ were grown on both the chip and the carrier. After being diced into smaller units, the chip and the carrier were then aligned and bonded together by a flip-chip bonder with a bonding force ranging from 5-10~kN. Before the bonding process, an important plasma cleaning process was performed to remove the oxides on the surface of the bumps.

In order to test the superconducting performance and yield rate of the indium bump connection, we first fabricated a series of test devices, which were designed as two braided wires, and the indium bumps acted as cross-connections~\cite{Foxen_2017}, shown in Fig.~\ref{fig:fab1} (b) insertion. Several pads were distributed at different points of the weaving to locate any connection failures. Tantalum films were first grown on two wafers by ultra-high vacuum DC magnetron sputtering, with the thickness of which was around 120 nm. Then the weaves, which were overlapped complementary arrays of bars, were patterned on the chip and the carrier by ultraviolet (UV) laser direct writing lithography (DWL). The Pads were also patterned for the current and voltage characteristics (I-V) measurement and arranged around the bar arrays on the carrier chip. After developing, the film were etched by reactive ion etching (RIE) with $\mathrm{SF_6}$. Then a total of 400 circular indium bumps were patterned on both sides of the overlap area by the DWL. After developing again, the indium was deposited by thermal evaporation. After being immersed in an acetone bath for several hours, the photo-resist was stripped and an array of bumps appeared.

The measured resistance-temperature (R-T) curves and the I-V characteristics of a test device in a physical property measurement system (PPMS) are shown in Fig.~\ref{fig:fab1} (a, b). Two obvious transitions are found in the R-T curves, corresponding to the superconducting transitions of tantalum ($\sim~4.3$~K) in $\alpha$ phase and indium ($\sim~3.4$~K), respectively. The I-V curve shows that the critical current of all bumps exceeds 10~mA. We do not reach the critical current due to the limitation of the current source. No failures are found in all batches of our test devices, which indicates that the yield of the indium bump fabrication process is almost 100\%.

The height of the indium bump directly determines the shunted capacitance of the flipmons. To precisely control the height of the indium bumps, we add some large indium bumps on the carrier as the stop bumps, as shown in Fig.~\ref{fig:measure} as large rectangle blocks. The stop bumps are fabricated simultaneously with the connection indium bumps. When the chip and the carrier are bonded together, the connecting bumps will first contact face-to-face and deform under the pressure until the stop bumps finally contact with the surface of the chip. Since the bonding force is constant, the pressure on each bump will drop sharply. The bumps deformation is almost ceased, and therefore the gap distance between two chips is determined by the height of  the stop bumps. The stop bumps can also help balance the force distribution on different chip areas so that the gap distance across the entire device is quite uniform. Fig.~\ref{fig:gap} shows the measured gap distances distribution of all nine samples and one of the relevant SEM images is shown in the inset. The gap distance can be well controlled in $5\pm 0.4\mathrm{\mu m}$. Since the capacitance of a parallel plate capacitor is inversely proportional to the gap distance, the distance fluctuation only causes less than 3\%'s variation of the qubit charging energy $E_c$. Then we extend the above fabrication process to our flipmon samples except for some extra steps to improve the quality. More detailed fabrication recipe of flipmon is shown in~Appendix.\ref{sec:fabrication}.

\section{\label{sec:meas} Measurement results}

In order to determine the stability of the fabrication process and the performance of the qubits, we have measured three flipmon samples with the same design. Fig.~\ref{fig:sample} shows an optical micrographic photo of a sample. Sample $\#1$ is wire-bonded in a multi-port sample holder. Sample $\#2$ and $\#3$ are wire-bonded in a two-port sample holder without using the XY control lines.


\begin{figure}[ht]
	\includegraphics{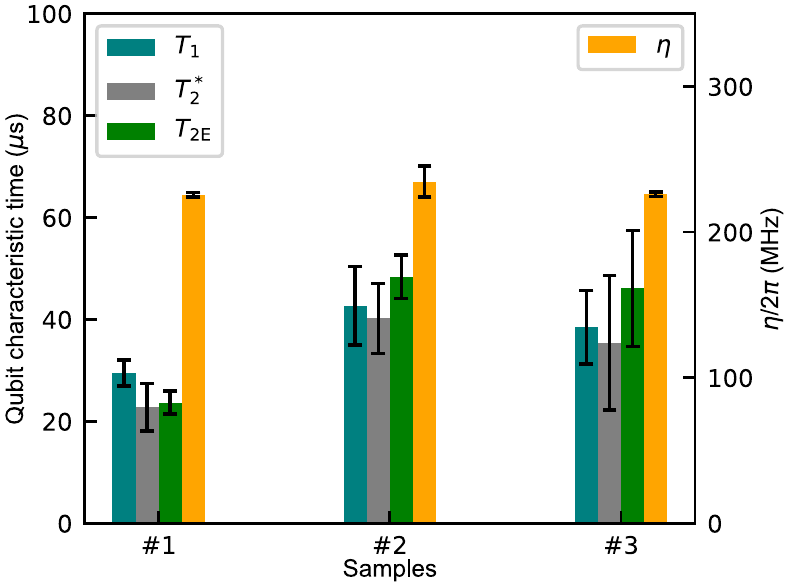}
	\caption{\label{fig:measure} The average of qubit characteristic times $T_1$, $T_\mathrm{2}^*$, $T_\mathrm{2E}$ and anharmonicity $\eta$ from three flipmon samples. Error bars are the fluctuations between different qubits in the same sample.}
\end{figure}

Each sample is well-packaged in a $\mu$-metal shield. We add adequate attenuation of 78~dB to the input transmission line to fully reduce the input thermal noise. The frequencies of the readout resonators are around 6.9~GHz, and the qubit frequencies are in the range of 4.5-5~GHz. The coherence characteristics of the three flipmon samples were measured, including energy relaxation time, decoherence time, and spin-echo time. The results are shown in Fig.~\ref{fig:measure}. Sample \#1 showed a lower coherence time, which may due to the poor packaging. Appendix \ref{sec:more-flipmon} provides detailed qubit characteristic measurement results of three qubit designs in Fig.~\ref{fig:design} (a), (b), (c).

The average energy relaxation time of flipmon sample \#2 and sample \#3 is in the range of 30-60 us. The sample designed as Fig.~\ref{fig:design} (b) are fabricated with the same recipe as flipmon and has almost the same energy relaxation time as flipmon, shown in Table~\ref{tab:barbell} and Table~\ref{tab:pplate}. This means that the indium bumps have little impact on the qubit relaxation time. However, in addition to the additional flip-chip process, the traditional transmons have a higher energy relaxation time in the range of 120-140 us using the same fabrication recipe as the flipmon. We believe the additional flip-chip process may not increase the loss tangent of the MS and SA interfaces. Therefore, the MA interfaces are most likely the dominant resource of flipmon energy relaxation due to the larger electric field participation ratio and the induced extra loss of the MA interface from the flip-chip process.


The sufficient anharmonicity $\eta$ is necessary to maintain transmon as a two-level state. Under the perturbation approximation, the anharmonicity is equal to the charging energy $E_c$ determined by the large shunt capacitor. Thus, we can measure the anharmonicity to infer the charging energy. We first pulse the qubit into the first excited state with a $\pi$ pulse and then scan the frequency between the first excited state and the second excited state. Then, we calculated the anharmonicity of three flipmon samples as shown in Fig.~\ref{fig:measure}. The almost same anharmonicity indicates that the vacuum gap distance determined by the indium bump height can be well-controlled due to the use of stop bumps.



\section{\label{sec:conclu}Conclusion}
In conclusion, we demonstrate a new design of transmon, denoted as flipmon, with the help of flip-chip technology. Good coherence properties that comparable to that of traditional transmons are obtained. The results show that our flip-chip bumps are of high quality and have little impact on the qubit energy relaxation. We have also developed a technology that uses the stop indium bumps to precisely control the gap distance between the chip and the carrier. The simulation of the electric field and the experimental results indicate the limitation of the energy relaxation time of the flipmons comes mainly from the MA interfaces. As a result, through the use of more careful surface treatment strategies~\cite{nersisyan2019manufacturing,tsioutsios2020free,mergenthaler2021ultrahigh}, the coherence of the flipmons can be further improved. Flipmons are naturally compatible with flip-chip technology, which is promising for the 3D wiring of qubit arrays. They are also compact and can provide flexible connectivity. We believe that flipmons and related quantum circuit architectures may be a good candidate for scaling quantum processors to the order of 1000 qubits or more.

\paragraph{}
\
\begin{acknowledgments}
This work was supported by the NSF of Beijing (Grant No. Z190012), the NSFC of China (Grants No. 11890704, No. 12004042,No. 11905100), National Key Research and Development Program of China (Grant No. 2016YFA0301800), and the Key-Area Research and Development Program of Guang Dong Province (Grant No. 2018B030326001).
\end{acknowledgments}

\appendix

\section{More Details of the samples}
\label{sec:sample image}
\begin{figure*}
	\includegraphics[width=.7\textwidth]{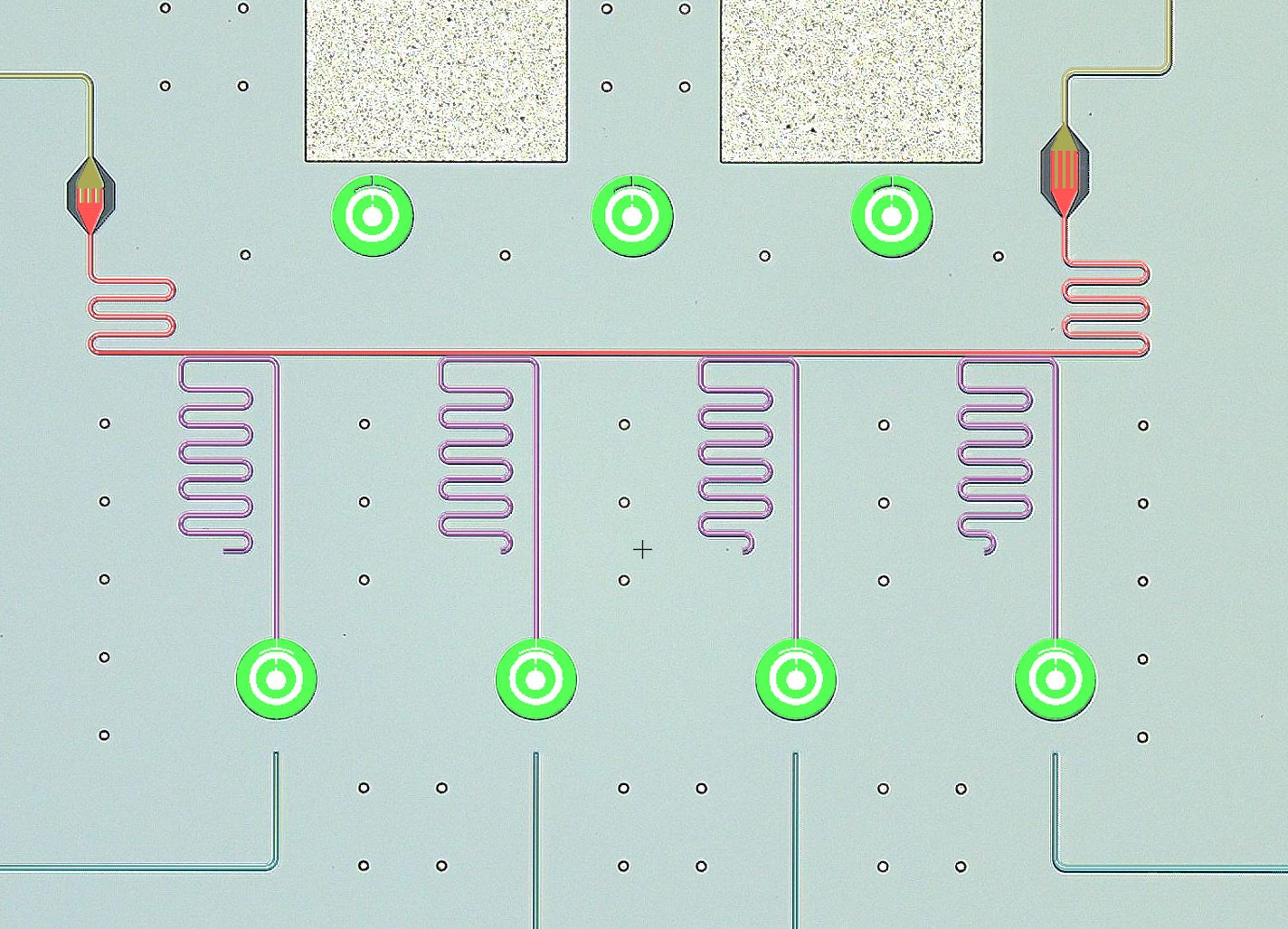}
	\caption{\label{fig:sample} Top view of a carrier containing four flipmons (green) with individual readout resonators (purple), three test flipmons for Josephson junction resistance measurement, one Purcell filter (red) embedded in one transmission line (yellow) and four XY control lines(cyan). Small circles are indium bumps used for connecting the carrier and the chip, while large rectangular bumps are the stop bumps.}
\end{figure*}

Fig.~\ref{fig:sample} shows an optical micrograph taken for the carrier chip of a flipmon sample before flip-chip bonding.
\label{sec:more-flipmon}
The measurement results of a traditional transmon sample are listed in Table~\ref{tab:transmon}. The qubit parameters of the sample designed as Fig.~\ref{fig:design} (b) are listed in Table~\ref{tab:pplate}. The parameters of each flipmon in Fig.~\ref{fig:measure} are listed in Table~\ref{tab:barbell}.
\begin{table*}[b]
	\caption{\label{tab:transmon}%
		Measured parameters of a traditional transmon sample.
	}
	\begin{ruledtabular}
		\begin{tabular}{lccccc}
			\textrm{ }&
			\textrm{Qubit1}&
			\textrm{Qubit2}&
			\textrm{Qubit3}&
			\textrm{Qubit4}&
			\textrm{Qubit5}\\
			\colrule
			Readout resonator frequency $f_\mathrm{r}$\,(GHz) & 7.101 & 7.133 & 7.166 & 7.198 & 7.226\\
			Transmon qubit frequency $f_\mathrm{q}$\,(GHz) & 4.450 & 4.559 & 4.412 & 4.418 & 4.502\\
			Anharmonicity $\eta/2\pi$\,(MHz) & --- & --- & 266 & --- & --- \\
			Dispersive shift $\chi/2\pi$\,(MHz) & 0.64 &0.71 & 0.65 & 0.63 & 0.65\\
			Relaxation time $T_1$\,($\mathrm{\mu}$s) & 158.3 & 109.2 & 136.3 &120.9 &131.2\\
			Ramsey dephasing time $T_\mathrm{2}^*$\,($\mathrm{\mu}$s) & 75.5 & 62.3 &42.0 & 52.1 &65.2\\
			Spin echo dephasing time $T_\mathrm{2E}$\,($\mathrm{\mu}$s) & 194.2 & 131.3 & 176.0 &173.8 &168.9\\			
		\end{tabular}
	\end{ruledtabular}
\end{table*}

\begin{table*}[b]
	\caption{\label{tab:pplate}%
		Measured qubit parameters of the sample designed as Fig.~\ref{fig:design} (b).
	}
	\begin{ruledtabular}
		\begin{tabular}{lcccc}
			\textrm{ }&
			\textrm{\#1Q1}&
			\textrm{\#1Q2}&
			\textrm{\#2Q1}&
			\textrm{\#3Q1}\\
			\colrule
			$f_\mathrm{r}$\,(GHz) & 6.625 & 6.725 & 6.917 & 7.190 \\
			$f_\mathrm{q}$\,(GHz) & 5.228 & 5.128 & 4.973 & 5.161 \\
			$\eta/2\pi$\,(MHz) & --- & 237 & 242 & ---\\
			$T_1$\,($\mathrm{\mu}$s) & 36.5 & 56.9 & 27.1 & 49.8 \\
			$T_\mathrm{2}^*$\,($\mathrm{\mu}$s) & 32.0 & 65.2 &29.9& 75.0 \\
			$T_\mathrm{2E}$\,($\mathrm{\mu}$s) & 73.0 & 95.1 & 30.2 & --- \\			
		\end{tabular}
	\end{ruledtabular}
\end{table*}

\begin{table*}[b]
	\caption{\label{tab:barbell}%
		Measured parameters of three flipmon samples. \#$i$Q$j$ is the $j$-th qubit on sample \#$i$.
	}
	\begin{ruledtabular}
		\begin{tabular}{lcccccccccccc}
			\textrm{ }&
			\textrm{\#1Q1}&\textrm{\#1Q2}&\textrm{\#2Q1}&
			\textrm{\#2Q2}&\textrm{\#2Q3}&\textrm{\#2Q4}&
			\textrm{\#3Q1}&\textrm{\#3Q2}&\textrm{\#3Q3}&
			\textrm{\#3Q4}&\textrm{\#3Q5}&\textrm{\#3Q6}\\
			\colrule
			$f_\mathrm{r}$\,(GHz) & 6.856 & 6.965 & 6.66 & 6.769 & 6.867 & 6.953 & 6.647 & 6.694 & 6.74 & 6.815 & 6.869 & 6.917\\
			$f_\mathrm{q}$\,(GHz) & 4.8 & 4.853 & 4.642 & 4.823 & 4.72 & 4.608 & 4.965 & 4.732 & 4.986 & 4.978 & 4.857 & 4.951\\
			$\eta/2\pi$\,(MHz) & 222.1 & 219.9 & 247 & 241.3 & 231.4 & 223 & 226.6 & 228.6 & 224.2 & 226.4 & 225 & 226.1\\
			$\chi/2\pi$\,(MHz) & --- & 0.32 & 0.47 & 0.44 & 0.41 & 0.3 & 0.5 & 0.4 & 0.8 & 0.6 & 0.6 & 0.4\\
			$T_1$\,($\mathrm{\mu}$s) & 27.7 & 31.3 & 43.7 & 39 & 53 & 35 & 33.8 & 32.6 & 47.5 & 30.1 & 44.4 & 42.7\\
			$T_\mathrm{2}^*$\,($\mathrm{\mu}$s) & 19.5 & 26.1 & 48.7 & 32.6 & 33.9 & 42.7 & 56.4 & 32.4 & 32.6 & 15.4 & 36.9 & 38.9\\
			$T_\mathrm{2E}$\,($\mathrm{\mu}$s) & 36.8 & 42.5 & 53.3 & 43.5 & 54.8 & 46.4 & 67.6 & 37.4 & 45.7 & 35.5 & 46 & 44.8\\			
		\end{tabular}
	\end{ruledtabular}
\end{table*}

\section{Details about the flipmon fabrication process}
\label{sec:fabrication}
In the fabrication process of flipmon, we need to keep in mind to improve the quality of the metal film. Special attention is paid to the removal of residual resist, which can otherwise be a primary source of dielectric loss \cite{niepce2020geometric}. Before the deposition of tantalum, the sapphire wafer was immersed in several inorganic solutions and annealed to keep the surface clean and flat. Then the Ta film with 120 nm was prepared by the dc magnetron sputtering. Next, all elements, except for the Josephson junctions, were patterned on the chip and the carrier by ultraviolet (UV) laser direct writing lithography (DWL). The inductively coupled plasma (ICP) etching with \ce{SF6} as etching gas was used to etch the tantalum. Notice, the oxygen plasma ashing was performed both before and after the ICP etching, for five and ten minutes, respectively, to remove the residual resist after the development and the resist-like chemicals deposited during etching.

The "T-shape" Josephson junctions (Fig.~\ref{fig:design}(f)) were patterned by electron beam lithography (EBL) on the carrier. The undercut structure and the Dolan bridge structure was constructed with PMMA A4/ LOR 10B bilayer resist. After two-minute oxygen plasma ashing, in-situ argon ion beam etching (IBE) was performed to remove the surface oxides of tantalum. Then the Josephson junction was formed by double-angle evaporation of aluminum. Next, the second DWL was used to pattern the indium bumps with a height of 5\,$\mathrm{\mu}$m to fit the designed vacuum gap distance. After the development again, we used the thermal evaporation to deposit the indium. Then the lift-off process was performed in the acetone, and followed by the immersion in N-Methylpyrrolidone (NMP) bath at 80$^{\circ}$C, which may be beneficial to the removal of residual resist around the junction region. A UV ozone treatment at 80$^{\circ}$C for ten minutes was helpful to remove the residual resist insoluble in the previous process thoroughly. The reducing gas treatment was extended to ten minutes to restore fresh surfaces of tantalum and indium. After quickly flip-chip bonding and wire bonding, we transferred the samples to the fridge as fast as possible to reduce the re-oxidation of metal surfaces.

\section{layout of flipmon 2D array}
\begin{figure}
	\includegraphics[width=.45\textwidth]{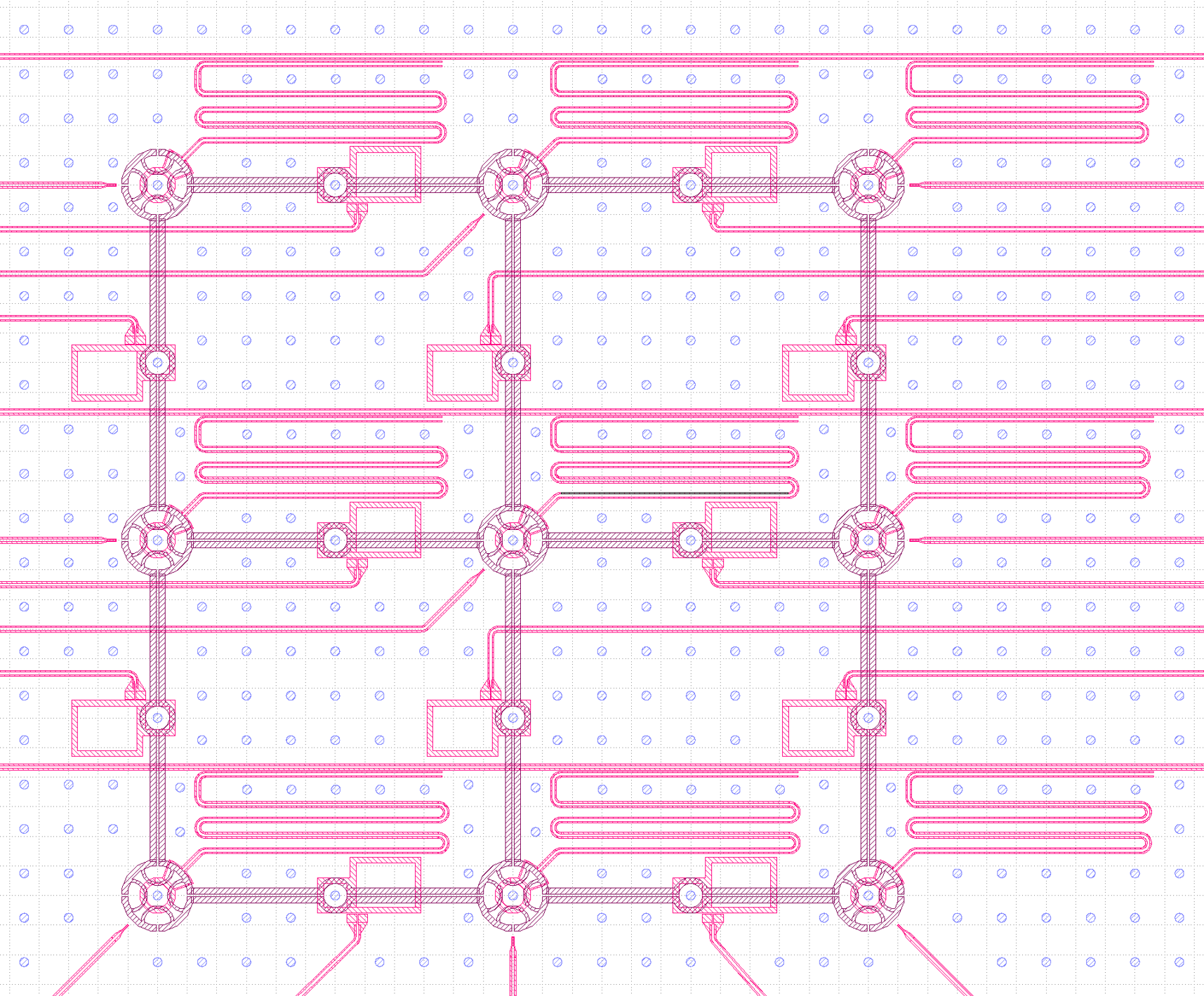}
		\caption{\label{fig:layout}Layout of a 2D square lattice of qubits including tunable couplers with the design of flipmon. The layer in magenta and plum color are metal to be etched on the carrier and the chip, respectively. The layer in blue is for indium bumps.}
\end{figure}

In order to demonstrate scalability, flipmon qubits are arranged in a two-dimensional square lattice, as shown in Fig.~\ref{fig:layout}. There is a $3\times3$ qubits array of flipmon with 12 tunable couplers. In the layout, qubits are a grounded version of flipmon, with the ring-shaped pad removed to avoid cutting up the ground plane. Tunable couplers are also the flipmon with rectangular pads, forming asymmetric floating couplers \cite{sete2021floating}. Readout resonators, transmission lines, control lines for direct current bias, and microwave driving are all on the carrier substrate. Without discontinuity caused by indium bumps on these lines, they may have better impedance matches and less loss or crosstalk. An array of indium bumps on both sides of coplanar lines balances the electric potential and suppresses spurious modes. Large rectangular bumps are stop bumps. Circular bumps on the capacitive shunt of qubits and couplers ensure that wiring between them can be put on the chip, separated in space with coplanar lines on the carrier. This may serve as an example of the flexibility in the wiring with the flipmon structure.

\label{sec:layout}

\providecommand{\noopsort}[1]{}\providecommand{\singleletter}[1]{#1}%

\end{document}